\documentclass[%
 aip,
 jmp,%
 amsmath,amssymb,
preprint,%
]{revtex4-1}

\usepackage{graphicx}
\usepackage{dcolumn}
\usepackage{bm}
\usepackage{colortbl}
\usepackage{setspace}

\begin{document}

\preprint{AIP/123-QED}

\title[UPOs in the generalized repressilator]{Transient dynamics around unstable periodic orbits in the generalized repressilator model}

\author{Natalja Strelkowa}
 \email{natalja.strelkowa06@imperial.ic.ac.uk}
\affiliation{Department of Bioengineering, Imperial College London, United Kingdom}
\author{Mauricio Barahona}
\email{m.barahona@imperial.ic.ac.uk}
\affiliation{Department of Bioengineering and Institute for Mathematical Sciences, Imperial College London, United Kingdom}

\date{\today}

\begin{abstract}
We study the spatio-temporal dynamics of the generalized repressilator, a system of coupled repressing genes arranged in a directed ring topology, and give analytical conditions for the emergence of a cascade of unstable periodic orbits (UPOs) that lead to reachable long-lived oscillating transients. Such transients dominate the finite time horizon dynamics that is relevant in confined, noisy environments such as bacterial cells (see our previous work {\it[Strelkowa and Barahona, 2010]}) and are therefore of interest for bioengineering and synthetic biology. We show that the family of unstable orbits possesses spatial symmetries and can also be understood in terms of traveling wave solutions of kink-like topological defects. The long-lived oscillatory transients correspond to the propagation of quasistable two-kink configurations that unravel over a long time. We also assess the similarities between the generalized repressilator model and other unidirectionally coupled electronic systems, such as magnetic flux gates, which have been implemented experimentally.
\\
\end{abstract}

\keywords{UPO | generalized repressilator model | genetic network | cyclic feedback | traveling waves}
\maketitle

\begin{quotation}
Bacterial cells are noisy environments that evolve over a limited time horizon. Under these conditions, the relevant dynamics is not dictated exclusively by the long-term attractors of the system but can be dominated by reachable long-lived transients with qualitatively different observable dynamics. Such transients are an important feature of naturally occurring biological networks, in particular in developmental biology, and can also be exploited for forward design of genetic elements in Synthetic Biology. Here, we investigate the dynamics of unidirectional rings of genetic repressors and show that their observable dynamics is dominated by oscillatory transients around unstable periodic orbits with strong spatio-temporal symmetries. The oscillatory transients around such orbits can be seen to correspond to the propagation of groups of kink-like excitations against the background of a dimerized solution, a picture with strong parallelisms with classic discrete lattice models~\cite{frenkelKont, Peierls55, Bulsara03a}, some of which have been implemented experimentally with electronic elements. 
\end{quotation}

\section{\label{sec:level1}Introduction}

The dynamics of symmetric interconnected systems are an intricate reflection of the inherent invariance associated with the symmetry present in the system~\cite{Golubitsky88}.  Engineered systems often resort to the use of symmetry because of the simplification in design and construction they afford, but also as a means to the production of controlled outcomes. Loops or ring structures of nonlinear elements~\cite{Yanchuk08, Landsman06, Perlikowski10} constitute one of the simplest topologies of practical interest in biology and engineering~\cite{Strogatz01, Bulsara04, Elowitz00, Wei99}.
For instance, in a recent series of papers ~\cite{Bulsara03a, Bulsara03b, Bulsara04, Bulsara06}  Bulsara {\it et al} reported the implementation of rings of unidirectionally coupled magnetic flux-gates through electronic circuits and their successful experimental use as highly sensitive magnetic field sensors~\cite{Bulsara03a, Bulsara03b, Bulsara04, Bulsara06}. The mathematical analysis of the steady-state dynamics of a variety of such ring systems has also revealed their rich dynamical behavior, including multi-stability, unstable periodic solutions, quasi-periodicity and even chaos and hyperchaos~\cite{Bulsara04, Bulsara06, Perlikowski10}.

These classic design concepts have found new applicability with the recent emergence of Synthetic Biology, a field which has put to use ideas from electrical engineering, computer science and physics to engineer circuits made up of biological parts, in many cases through the design of gene regulatory networks~\cite{Heinemann06, Khalil10}. Such synthetic circuits of interconnected genes have been implemented in living microbes to perform low-level functions, such as logical gates~\cite{Ramalingam09}, switches~\cite{Gardner00} and genetic oscillators~\cite{Elowitz00},  exploiting the steady-state behavior of the gene network. 

\begin{figure}
 \includegraphics[width=6.4in]{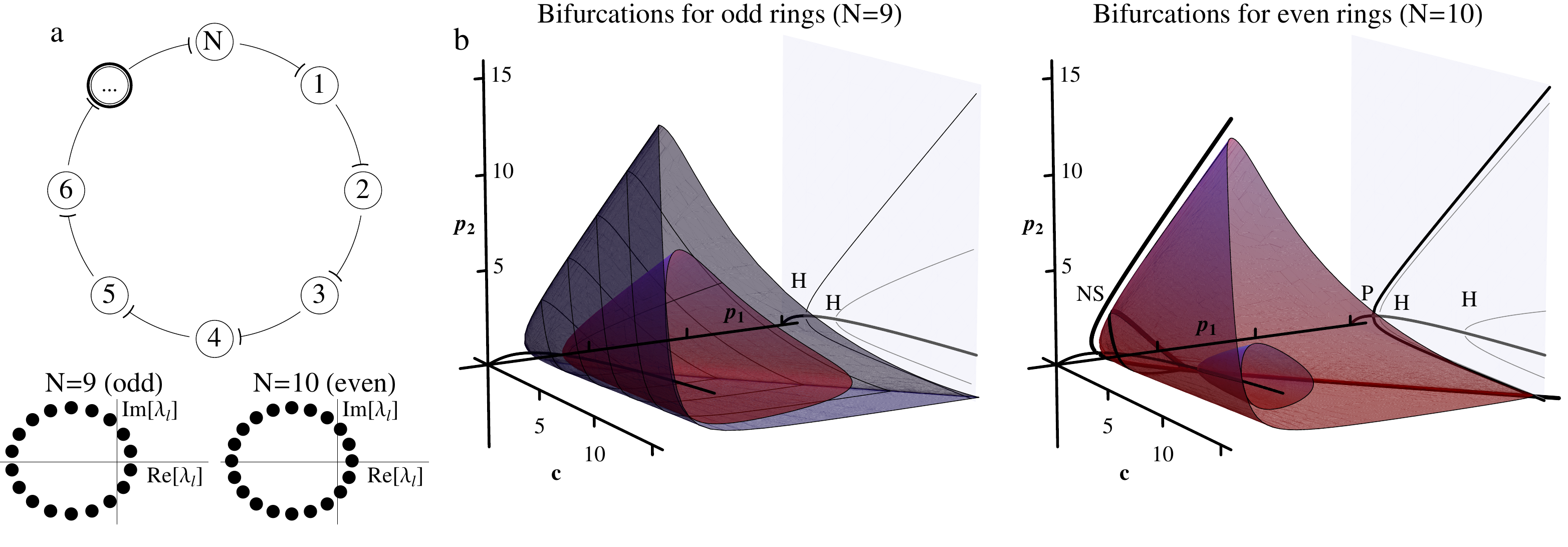}
  \caption{ {\bf The generalized repressilator model} {\bf (a)} The network consists of genes interconnected in a directed ring topology, in which the protein product of gene $j$ represses the expression of gene $j+1$ with periodic boundary conditions.
{\bf (b)}  As the parameter $c$ (see Eq.~\ref{eq:critical}) is increased, a cascade of Hopf bifurcations emerges---here shown through $p_1$ vs. $p_2$ phase plots. In odd rings (exemplified by $N=9$), an attracting limit cycle (meshed style) emerges from the first Hopf bifurcation (H). In even rings (exemplified by $N=10$), a pitchfork bifurcation (P) leads to two symmetrical stable fixed points (solid black lines), which we refer to as `up/down' solutions. In both even and odd rings, further Hopf bifurcation lead to emergence of UPOs.  The difference in the bifurcation cascades between even and odd gene rings can be understood from the pattern of the eigenvalues of the linearization around the symmetric fixed point $p_m$.  Some limit cycles in even rings undergo Neimark-Sacker bifurcations (NS) leading to an unstable torus.}
  \label{fig:geneRing}
  \end{figure}

Here, we study the spatio-temporal dynamics of the generalized repressilator, 
a gene regulatory network of repressing genes cyclically coupled in a directed 
ring topology (see Fig.~\ref{fig:geneRing}a), 
and we concentrate on the investigation of the role of unstable periodic orbits (UPOs) on the observable transient dynamics. 
The study of the generalized repressilator model has a long history~\cite{Fraser74,Smith87,Mueller06}. 
As shown analytically in~\onlinecite{Smith87}, there is a fundamental difference between rings with odd and even number of monotonic repressors: at steady state, odd-numbered rings possess a limit cycle attractor while even-numbered rings are bistable and act as switches. Hence the generalized repressilator model bridges between switches and oscillators in a natural way. In fact, the first ever genetic oscillator realized in a living bacterium by \citet{Elowitz00} is a special case of the generalized repressilator for three genes, while the first genetic switch implemented in E. coli by \citet{Gardner00} corresponds to the generalized repressilator with two genes. 

Previous mathematical work on these systems was mainly devoted to stability analysis and the characterization of the attractors that determine the long term behavior of the rings. On the other hand, the study of the transient behavior, in particular in relation to the presence of UPOs, was broadly ignored. However, the standard steady-state analysis alone is not sufficient in systems where transients are qualitatively different from steady states and extend over
long times compared to the observational time scales of the system, e.g., the average life time of a bacterial cell for a system like the repressilator. 
Even numbered repressilator rings are examples of such systems: although their steady state is typically bistable, with two attracting fixed points~\cite{Smith87}, the fixed points are approached through long-lived oscillating transients that dominate the observable dynamics~\cite{Strelkowa10}. In a recent paper, we showed that this transient behavior could be exploited to produce switchable oscillations by using feedback to operate the system around these unstable oscillations~\cite{Strelkowa10}. 

In this paper, we show that the observed long-lived transients have their origin in the existence of UPOs that attract initial conditions to their vicinity, thus funneling the dynamics into a lower-dimensional oscillatory subspace from which it is difficult to escape.  The UPOs appear through a cascade of Hopf bifurcations, for which we provide analytical conditions, and possess spatio-temporal symmetries consistent with tilings of the ring with smaller basic solutions~\cite{Barahona97}. In addition, the dynamics around the UPOs can also be understood in terms of traveling kink-like excitations propagating against the backdrop of a dimerized discrete lattice---a picture analogue to other propagating waves in classical physics~\cite{Peierls55,frenkelKont, Altland06}. The reachability and persistence of the dynamics around the UPOs, which make them observable, is related to the interaction and propagation of moving kinks in the lattice. Finally, we show that these features also apply to the analysis of rings of magnetic flux-gates~\cite{Bulsara03a} implemented experimentally with electronics---a system in which extremely long-lived oscillating transients (especially in the presence of noise) have been observed both experimentally and numerically in even-numbered rings that should be bistable according to steady state analysis.

\section{Bifurcation analysis of the generalized repressilator and UPOs}
%
A standard model of the generalized repressilator with $N$ genes is given by the following ODEs:
\begin{eqnarray}
\dot{m}_j&=&c_1 f(p_{j-1})-c_2 m_j \nonumber\\
\dot{p}_j&=&c_3m_j-c_4p_j \quad\quad\quad j=1,\dots,N.
\label{eq:system}
\end{eqnarray}\
Here, $p_j$ and $m_j$ describe protein and mRNA concentrations for each gene (with periodic boundary conditions such that $p_0\equiv p_N$). The kinetic constants $c_1$ and $c_3$ represent the strength of repression and transcription, respectively, while $c_2$ and $c_4$ correspond to degradation of each species.  The repressing behavior is encapsulated by $f(p)$, a positive function $f:\mathbb{R}^+\rightarrow\mathbb{R}^+$ that is bounded between 0 and 1 and is monotonically decreasing over its domain. For such a function, the system~(\ref{eq:system}) has a spatially uniform fixed point~\cite{Smith87} in which all proteins have the same concentration: $p_j=p_{j-1}=p_m \; \forall j$.

Throughout this paper, our numerics are performed using a standard Hill-type function, which is well
established in biochemical modeling: 
\begin{equation}
f(p) = \frac{1}{1+p^h},
\label{eq:Hill}
\end{equation}
where $h$ is the Hill coefficient. However, the outlined stability analysis is similar for any positive function that is bounded and monotonically decreasing. For the Hill function~(\ref{eq:Hill}), the uniform fixed point solution $p_m$ follows from solving the polynomial equation:
\begin{eqnarray}
p_m+p_m^{h+1}= \frac{c_1c_3}{c_2c_4} \equiv c.
\label{eq:critical}
\end{eqnarray}
For the case $h=2$, which is relevant in some Synthetic Biology implementations~\cite{Elowitz00}, the cubic equation that ensues can be solved explicitly using the Cardano method. Note that this solution branch is independent of the number of genes in the ring and exists for all positive $c$. In fact, $p_m$ is the only fixed point when the repression (or coupling) $c$ is small enough.

As the strength of the repressive coupling $c$ increases, there is a qualitative difference between the bifurcations undergone by even- and odd-numbered rings (see Fig.~\ref{fig:geneRing}). In odd rings, 
$p_m$ becomes unstable via a Hopf bifurcation resulting in a stable limit cycle. Even rings, on the other hand, first undergo a pitchfork bifurcation where two branches of stable dimerized solutions emerge. No further stable solutions appear as $c$ is increased either for even or odd rings. Hence from the viewpoint of strict stability analysis, odd rings behave as oscillators, with a globally attracting limit cycle, and even rings behave as switches, due to their bistability. 

There are however other bifurcations that lead to the emergence of unstable solutions. In both even and odd rings, further increasing the coupling $c$ leads to a cascade of Hopf bifurcations, where each bifurcation gives rise to a UPO that emerges from the already unstable branch $p_m$. In addition, numerical continuation~\cite{MATCONT}  also reveals the existence of Neimark-Sacker bifurcations in particular UPOs of even rings, in which an unstable torus emerges from the UPO. Figure~\ref{fig:geneRing} shows characteristic examples of these behaviors. 

The conditions for the appearance of UPOs can be derived analytically from the instabilities of the spatially
uniform fixed point $p_m$. Eliminating the mRNA variables ($m_j$), Eq.~(\ref{eq:system}) can be rewritten as:
\begin{eqnarray}
\ddot{p}_j&=& c_1 c_3 f(p_{j-1}) -(c_2+c_4)\dot{p}_j - c_2c_4p_j.
\label{eq:protein_second_order}
\end{eqnarray}
This corresponds to the system
\begin{eqnarray}
\dot{p}_j&=&q_j\nonumber\\
\dot{q}_j&=&c_1 c_3 f(p_{j-1}) -(c_2+c_4)q_j - c_2c_4p_j,
\label{eq:protein_first_order}
\end{eqnarray}
with linear and nonlinear parts as follows:
\begin{equation}
{\bf \dot z}=(I_N \otimes A) {\bf z} + H({\bf z}).
\label{eq:vectorsystem}
\end{equation}
Here, ${\bf z}^T = \left(\ldots  p_j \, q_j \ldots \right)$, $I_N$ is the $N$-dimensional identity matrix,  
 $A = \small{\left( \begin{array}{cc}
0 & 1 \\ 
-c_2c_4 & -(c_2+c_4) \\
\end{array} \right)}$
and $H({\bf z})$ is a vector that contains the nonlinear coupling between the elements. 

Linearizing around the fixed point $p_m$ we obtain the Jacobian 
\begin{equation}
J|_{p_m}=(I_N \otimes A) + (G \otimes H^*)
\label{eq:jacobian}
\end{equation}
where $H^*= c_1 c_3 \left( \begin{array}{cc}
\small{0} & \small{0} \\
\left .{\frac{df}{dp}}\right |_{p_m}& \small{0} \\
\end{array} \right)$ and the directed ring topology of the network is encoded in 
the $N \times N$ adjacency matrix { \small \[  G = \left( \begin{array}{ccccc}
0 & 0 &0 &\dots & 1 \\ 
1 & 0 &0 & \dots &0 \\
. & . &. &. & . \\
0 & \dots &1 &0 &0 \\
0 & 0 & \dots &1 &0 
\end{array} \right). \] }
Fourier-diagonalizing~\cite{Barahona02} the circulant matrix $G$, 
we obtain the following complex quadratic equation for the eigenvalues $\lambda_\ell$ of the 
Jacobian~(\ref{eq:jacobian}): 
\begin{equation}
\lambda_\ell^2 + (c_2 +c_4) \lambda_\ell + c_2c_4\left(1- c \left .{\frac{df}{dp}}\right |_{p_m} e^{i \frac{2 \pi}{N} \ell} \right)=0
\label{eq:characteristic_polynomial}
\end{equation}
with $\ell=\{0,..,N-1\}$. 

The necessary condition for a Hopf bifurcation, i.e., the existence of purely imaginary roots, leads to inequalities on the parameters~\cite{Hardy} that determine the number of bifurcations. It then
follows that the maximum number of Hopf bifurcation is $(N/2-1)$ for even rings and $(N-1)/2$ for odd rings (see Appendix). 

If the repressor is described by the Hill function~(\ref{eq:Hill}), we have 
$$\left .{\frac{df}{dp}}\right |_{p_m} =  -\frac{h p^{h+1}_m}{c^2}.$$  When $h=2$ and with parameters relevant for the reported Synthetic Biology experiments~\cite{Elowitz00},
 the number of Hopf bifurcations is limited to 
$ \left\lfloor\frac{1}{2}\frac{(N+1)}{2}\right\rfloor$ for odd rings and 
$\left \lfloor \frac{1}{2} \left(\frac{N}{2}-1\right)\right\rfloor$ for even rings (see Appendix and Table~\ref{tab:UPOSummary}). 

Apart from the first bifurcation in odd rings, which gives rise to an attracting limit cycle, all other Hopf bifurcations give rise to unstable periodic orbits (UPOs), as shown in Figure~\ref{fig:geneRing} and summarized in Table~\ref{tab:UPOSummary}. Although UPOs do not lead to long term attractors, they can still be important for practical applications if they are likely to be reached from random initial conditions (or as a result of random fluctuations) and they dictate long-lived dynamics that will therefore feature in finite-time observations. As shown below, the UPOs that emerge from the first Hopf bifurcation in even rings have these characteristics for parameter ranges consistent with biological experiments. In particular, with mRNA and protein life-times on the order of minutes and Hill coeficients $h \in \{2,\dots,6\}$, the transient oscillatory behavior would extend over several hours in rings of moderate size $N \sim 10$ when the cellular doubling time is on the order of one hour. Therefore the transient behavior dominates the observable dynamics. 

We now explore the different dynamical facets of the emerging unstable periodic orbits. In particular, we describe their spatio-temporal symmetry, and their relationship to tilings of the ring, and present their linear stability. We also show that these solutions can be understood in terms of traveling waves that correspond to propagating interacting kinks, analogous to solutions from classical physics such as domain walls.  We also show that,  although unstable, these periodic solutions are reachable from random initial conditions and long-lived and, as such, they are relevant in finite-time dynamics in real-world applications.

\section{Spatio-temporal symmetries of UPOs of the generalized repressilator model}
\begin{figure}[h!]
	\includegraphics[width=0.75\textwidth]{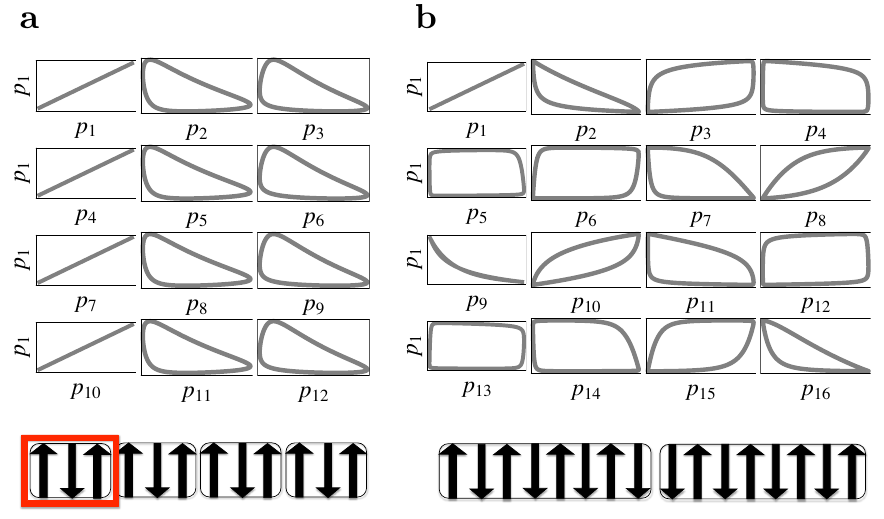}
\caption{ {\bf Spatio-temporal symmetries in the generalized repressilator model}. \textbf{(a)} Phase plots for the symmetric UPO originating from the second Hopf bifurcation in the 12-gene ring. We plot the time evolution of the first gene against other genes in the ring. Genes that are dynamically equivalently are plotted column-wise. Each of the four rows shows a group of genes, which behave dynamically in the same way as the 3-gene ring. The solution has four kinks corresponding to the domain walls between the 3-gene tiles, as shown in the bottom diagram. \textbf{(b)} Phase plots for the non-symmetric UPO originating from the first Hopf bifurcation in the 16-gene ring.  In this case there is no obvious symmetry between the different genes. As shown in the bottom diagram, this a basic solution that contains two kinks.}
\label{fig:bifSum}
\end{figure}

The discrete rotational symmetry of the equations~(\ref{eq:system})  induces non-trivial symmetries in the UPOs which can be directly related to the linear stability of the solutions.  Table~\ref{tab:UPOSummary} presents a summary of the bifurcation analysis and emerging solutions for $N \in [3,18]$ carried out with the continuation software MATCONT~\cite{MATCONT}. The analysis reveals that some of the higher order bifurcations for larger rings have the same periods as lower order bifurcations in smaller rings. This is the result of a rotational symmetry of such solutions that correspond to a tiling of the ring, i.e., gene $p_j$ has exactly the same time behavior as gene $p_{j+s}$. 
This can be seen, for instance, in the unstable periodic solution that emerges from the second Hopf bifurcation in the $N=12$ gene ring, where $p_j, p_{j+3}, p_{j+6}$ and $ p_{j+9}$ have the same time traces (see Fig.~\ref{fig:bifSum} a). Each triplet of consecutive elements follows a dynamics equivalent to that of a 3-gene ring, i.e., the solution corresponds to the tiling of the 12-gene ring by four 3-gene repressilators.  Therefore, its period is the same as that of the 3-gene ring for the same set of constants.

Table~\ref{tab:UPOSummary} summarizes the Hopf bifurcations and some of the properties of the ensuing periodic solutions (limit cycles and UPOs) obtained through numerical continuation and from the analysis of numerical simulations. We have considered rings of length between $N=3$ and $18$ of the model with Hill-type repression with $h=2$ for a range of the bifurcation parameter $c_1\in(0,3]$. The observed bifurcations follows the number and pattern of Hopf bifurcations found analytically (see Appendix). The solutions are sorted by increasing period and therefore symmetric tiling solutions appear grouped together (since they have the same period). Not all the solutions correspond to tilings of the ring. Some of the UPOs are `basic solutions' with no strict rotational symmetry, i.e., no two genes behave in the same way (Figure~\ref{fig:bifSum}b). The fact that some solutions of the system can be understood in terms of tilings of basic solutions indicates the importance of domain walls and kink-like excitations in the understanding of the observed solutions. We explore this alternative dynamical picture below. 

\begin{singlespace} 
\begin{table}
\begin{tabular}{|c|c|c|c|c|c|}
\hline
N  & 
$\begin{array}{c}
 \text{Unstable}\\
 \text{Floquet}
 \end{array}$       & 
 Period  &
  $\begin{array}{c}
 \text{No. of}\\
 \text{kinks}
 \end{array}$  & 
 $\begin{array}{c}
 \text{Basic solution}\\
 \text{(kinks)}
 \end{array}$ &
 $\begin{array}{c}
 \text{No. of}\\
 \text{tilings}
 \end{array}$\\
\hline
11 &        4/22&	86  &         5     &11(5) & 1\\
18 &	7/36&	89  &         8     &18(8)& 1\\
\rowcolor[rgb]{0.,0.7,0.7}
7   &	2/14&	94  &         3      &7(3)& 1\\
\rowcolor[rgb]{0.,0.7,0.7}
14 &	5/28&	94  &         6	& 7(3)& 2\\
17 &	6/34&	100&         7	&17(7)&1\\
10 &	3/20&	104&	 4	&10(4)&1\\
13 &	4/26&	110&	 5	&13(5)&1\\
16 &	5/32&	113&	 6	&16(6)& 1\\
\rowcolor[gray]{.8}
\textcolor[rgb]{1,0,0}{\bf{3}}&\textcolor[rgb]{1,0,0}{\bf{0/6}}&
\textcolor[rgb]{1,0,0}{\bf{132}}&\textcolor[rgb]{1,0,0}{\bf{1}}&
\textcolor[rgb]{1,0,0}{ \bf{3(1)}}&\textcolor[rgb]{1,0,0}{\bf{1}}\\
\rowcolor[gray]{.8}
6   &	1/12&	132&	 2	&3(1)& 2\\
\rowcolor[gray]{.8}
9   &	2/18&    132&	 3	&3(1)& 3\\
\rowcolor[gray]{.8}
$12^\ast$ &	3/24&132&	 4	&3(1)& 4\\
\rowcolor[gray]{.8}
15 &        4/30&	132&	 5 	&3(1)& 5\\
\rowcolor[gray]{.8}
18 &	5/36&	132&	 6	&3(1)& 6\\
17 &	4/34&	154&	 5	&17(5)& 1\\
14 &	3/28&	160&	 4	&14(4)& 1\\
11 &	2/22&	169&	 3	&11(3)& 1\\
\rowcolor[rgb]{0.5,0.5,1}
8   &	1/16 &	186&	 2	&8(2)& 1\\
\rowcolor[rgb]{0.5,0.5,1}
16 &	3/32&	186&	 4	&8(2)& 2\\
13 &	2/26&	204&	 3	&13(3)& 1\\
18 &	3/36&	213&	 4 	&18(4)& 1\\
\rowcolor[rgb]{0.85,0.3,0.7}
5   &	0/10&	239&	 1	&5(1)&1\\
\rowcolor[rgb]{0.85,0.3,0.7}
10 &	1/20&	239&	 2	&5(1)&2\\
\rowcolor[rgb]{0.85,0.3,0.7}
15 &	2/30&	239&	 3	&5(1)&3\\
17 &	2/34&	274&	 3	&17(3)&1\\
12 &	1/24&	291&	 2	&12(2)& 1\\
\rowcolor[rgb]{1,0.5,0}
7   &	0/14&	342&	 1	&7(1)&1\\
\rowcolor[rgb]{1,0.5,0}
14 &	1/28&	342&	 2	&7(1)&2\\
$16^\ast$ &	1/32&	393&	 2	&16(2)&1\\
\rowcolor[rgb]{1,1,0}
9   &	0/18&	444&	 1      &9(1)&1\\
\rowcolor[rgb]{1,1,0}
18 &	1/36&	444&	 2	&9(1)&2\\
11 &	0/22&	545&	 1	&11(1)&1\\
13 &	0/26&	645&	 1	&13(1)&1\\
15 &	0/30&	746&	 1	&15(1)&1\\
17 &	0/34&	847&	 1	&17(1)&1\\
\hline
\end{tabular}
\caption{{\bf Bifurcations, symmetries and kinks in the generalized repressilator model.} The analytical conditions (Eq.~\ref{eq:characteristic_polynomial}) for the emergence of UPOs are confirmed  for rings of length $N=\{3,\ldots,18\}$ through simulations and numerical continuation~\cite{MATCONT}. The numerics have been carried out for the Hill function with  $h=2$ for the following set of parameters: $c_2 = 0.12\,\text{min}^{-1}, c_3 = 0.16\,\text{min}^{-1}, c_4 = 0.06\,\text{min}^{-1}$, and the bifurcation parameter $c_1$ in the range $(0, 3]$. We have followed the UPOs numerically calculating their Floquet exponents, period, and spatio-temporal symmetry. The number of kinks has been numerically extracted from time traces according to Eq.~(\ref{eq:kinkTrafo}). The UPOs are sorted by increasing period and solutions whose dynamics are equivalent by tiling are highlighted with the same color. For instance, the basic 3-gene repressilator (bold red text) has an associated family of symmetric tiling solutions for rings where the gene number $N$ is a multiple of three. Phase plots of the two solutions marked with asterisks are presented in Figure~\ref{fig:bifSum}. }
\label{tab:UPOSummary} 
\end{table}
\end{singlespace}

\section{UPOs as traveling kink solutions}
An alternative viewpoint to understand the dynamics of gene rings is to view the periodic solutions in terms of propagating kinks against the backdrop of a dimerized (`up-down') solution. Analogues to this picture are abundant in both the nonlinear waves literature and in classical examples in condensed matter physics, such as propagation of domain walls in antiferromagnetic chains or soliton-like excitations in dimerized polymer chains~\cite{Peierls55,frenkelKont,Altland06}.

The link between the dynamical bifurcations presented above and the kink-domain picture follows straightforwardly from the examination of the fixed points of the system. As explained above, the uniform solution, where all proteins have the same value $p_j=p_m, \, \forall j$, is always a fixed point of the system and is the global attractor of the equations for small enough values of the coupling parameter $c$.   

When the coupling between the genes is large enough, the uniform solution becomes unstable either through a Hopf bifurcation (for odd rings) or through a pitchfork bifurcation (for even rings).  The two stable fixed points that emerge from the pitchfork bifurcation in even rings correspond to dimerized solutions (`up-down') where genes are over-expressed ($p_u  \textgreater p_m$) or under-expressed ($p_d \textless p_m$) in an alternating sequence around the ring. This arrangement leads to two degenerate solutions, equivalent by a one-gene rotation,  which correspond to the sequences `up-down' and `down-up', respectively. Clearly, such a dimerized solution is compatible with an even ring but not with an odd ring. 

\begin{figure}[h!]
	\includegraphics[width=0.75\textwidth]{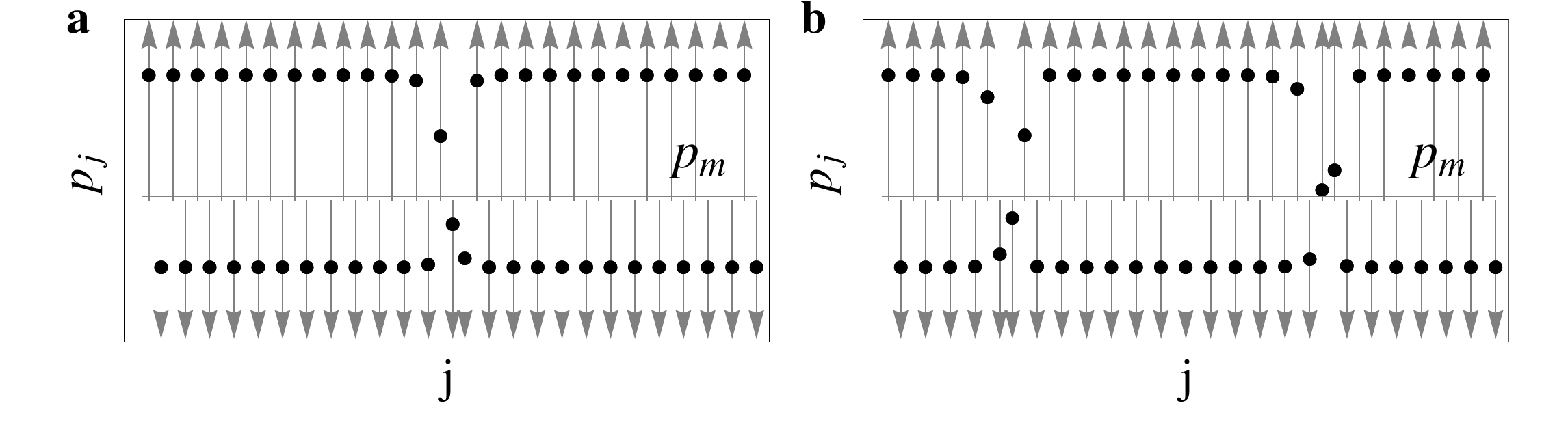}
  \caption{ {\bf Kinks in the repressilator model. } The kink-domain picture provides an alternative characterization of periodic orbits. Numerical time snapshots in gene space (black dots) are mapped onto domains and kinks using Eq.~\ref{eq:kinkTrafo}  (gray arrows).  {\bf (a)} Frustrated solution in an odd-numbered (N=51) gene ring, which emerges from the first Hopf bifurcation and is a stable attractive limit cycle. Its stability follows intuitively from the fact that the kink can propagate along the ring forever. {\bf (b)} The solution with two kinks in the even-numbered ring (N=50) is, on the other hand, unstable. This solution originates from the first Hopf bifurcation in an even ring. The dynamics around this UPO corresponds to the motion of the two kinks along the ring with a slightly different speed. As one kink gains on the other, they eventually annihilate and the ring becomes quiescent. Importantly for applications, this happens only after a very large number of periods ($\sim 100$).
}
\label{fig:kink50}
\end{figure}

In the case of an odd ring, trying to fit an `up-down' solution leads to frustration due to the boundary conditions, originating a \textit{kink} mismatch when wrapping up the sequence.  Indeed, the first Hopf bifurcation in odd rings can be viewed as the emergence of a propagating solution of this type. Hence the attracting limit cycles in odd rings correspond to the fundamental solution with one traveling kink propagating against a background of the dimerized solution. The localized structure of the kink (domain wall) is shown in Figure~\ref{fig:kink50}a for a  ring of size $N=51$. A quantitative description of the dynamics of the kinks can be obtained from numerical simulations by defining the threshold function:
\begin{eqnarray}
 \nu(p_j) = 
  \begin{cases} 
   1 & \text{if } p_j \geq p_m  \quad \equiv \text{`up'} \\
   -1 & \text{if } p_j  < p_m \quad \equiv \text{`down'} ,
  \end{cases}
\label{eq:kinkTrafo}
\end{eqnarray}
which allows us to define the domain regions as sequences of alternating genes. The location of the kinks is
then defined by the condition $\nu(p_j) =  \nu(p_{j+1})$.

The pattern now becomes clear. The cascade of Hopf bifurcations corresponds to the emergence of solutions with an increasing number of kinks in the ring. The first Hopf bifurcation in \textit{even} rings corresponds to the emergence of a solution with two traveling kinks which can be seen as localized excitations in the ring, as seen in  Figure~\ref{fig:kink50}b for a  ring of size $N=50$.
The kinks (or walls) originate from the mis-matching of two 'up-down' domains to give an arrangement that can be fitted into an even ring, as shown at the bottom of Figure~\ref{fig:bifSum}b. As indicated by Floquet analysis, this solution is an unstable periodic orbit in which the two kinks get closer as time evolves until they annihilate each other. As the two kinks conflate, the domain separation unravels and we recover the 'up-down' dimerized solution.  Similarly, the second Hopf in \textit{odd} rings gives rise to UPOs that correspond to solutions containing three traveling kinks, while the second Hopf in \textit{even} rings  leads to the emergence of UPOs with four propagating kinks in the ring. 

\begin{figure}[h!]
\begin{center}
\includegraphics[width=0.65\textwidth]{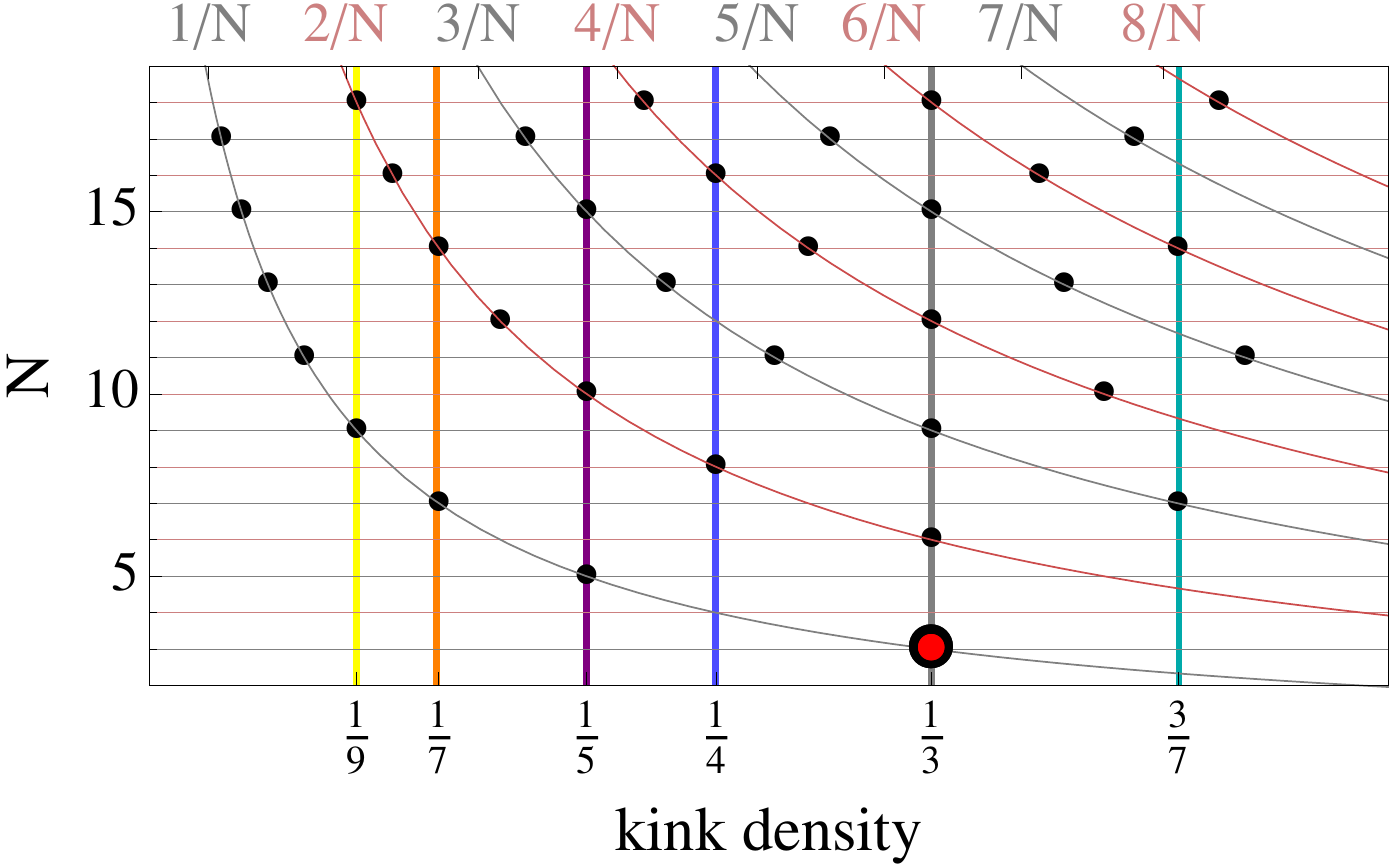}
\caption{ {\bf Tilings and kink density of UPOs of the generalized repressilator model}. Summary of the
solutions presented in Table~\ref{tab:UPOSummary} with the same parameters. 
For this set of parameters the kink density is always smaller than $1/2$. Solutions
with the same number of kinks are connected by the grey (odd) and pink (even) curves.  
The solutions of odd-numbered rings appear at the crossings of the grey horizontal lines with the grey curves. 
Similarly, the intersections of the pink horizontal lines and the pink curves mark the UPOs for even rings. 
The colored vertical lines connect the families of symmetry tilings (colors matched to those in 
Table~\ref{tab:UPOSummary}). 
}
\label{fig:kinkDens}
\end{center}
\end{figure}

It then follows that even numbered rings only admit (unstable) configurations with an even number of kinks that eventually annihilate to recover the stable dimerized solution. In odd rings, only solutions with an odd number of kinks are possible and, similarly, all the kinks but one eventually annihilate, leaving a single kink propagating along the ring indefinitely, i.e., we recover the stable limit cycle basic solution. Note also that when the solutions correspond to tilings of basic solutions, the total number of kinks is the result of multiplying the number of kinks of the basic solution by the number of tilings. Moreover, the number of Hopf bifurcations, which follows from symmetry constraints,  is therefore linked to the maximum number of kinks that can be accommodated in the rings. A summary of the periodic solutions in Table~\ref{tab:UPOSummary} is presented  in terms of their kink density in Figure~\ref{fig:kinkDens}.

\section{Observable long-lived oscillating transients around UPOs}

The kink-domain picture not only provides insight into the existence and structure of periodic solutions, 
but also gives an intuitive understanding of their stability. As expected, the larger the number of kinks, the more unstable the solutions become. This can be seen in the fact that the number of unstable Floquet directions grows as the number of kinks increases: the number of unstable Floquet directions is equal to the number of kinks minus one (Table~\ref{tab:UPOSummary}). In addition, the increased instability of the solutions with a large number of kinks also translates into a reduced length of the transients observed numerically, i.e., the time it takes for the kinks to conflate decreases as the density of kinks increases. This means that for practical purposes, the dominant long-lived oscillatory transients will be associated with the UPO emerging from the first Hopf bifurcation in even rings. We show below that the length of the oscillatory transients is long compared to the observation time scales of the system, hence these UPOs dictate the observable dynamics of the system.

\subsection*{Slow convergence to the dimerized solution through kink-kink annihilation}
\begin{figure}[ht!]
 \includegraphics[width=.6\textwidth]{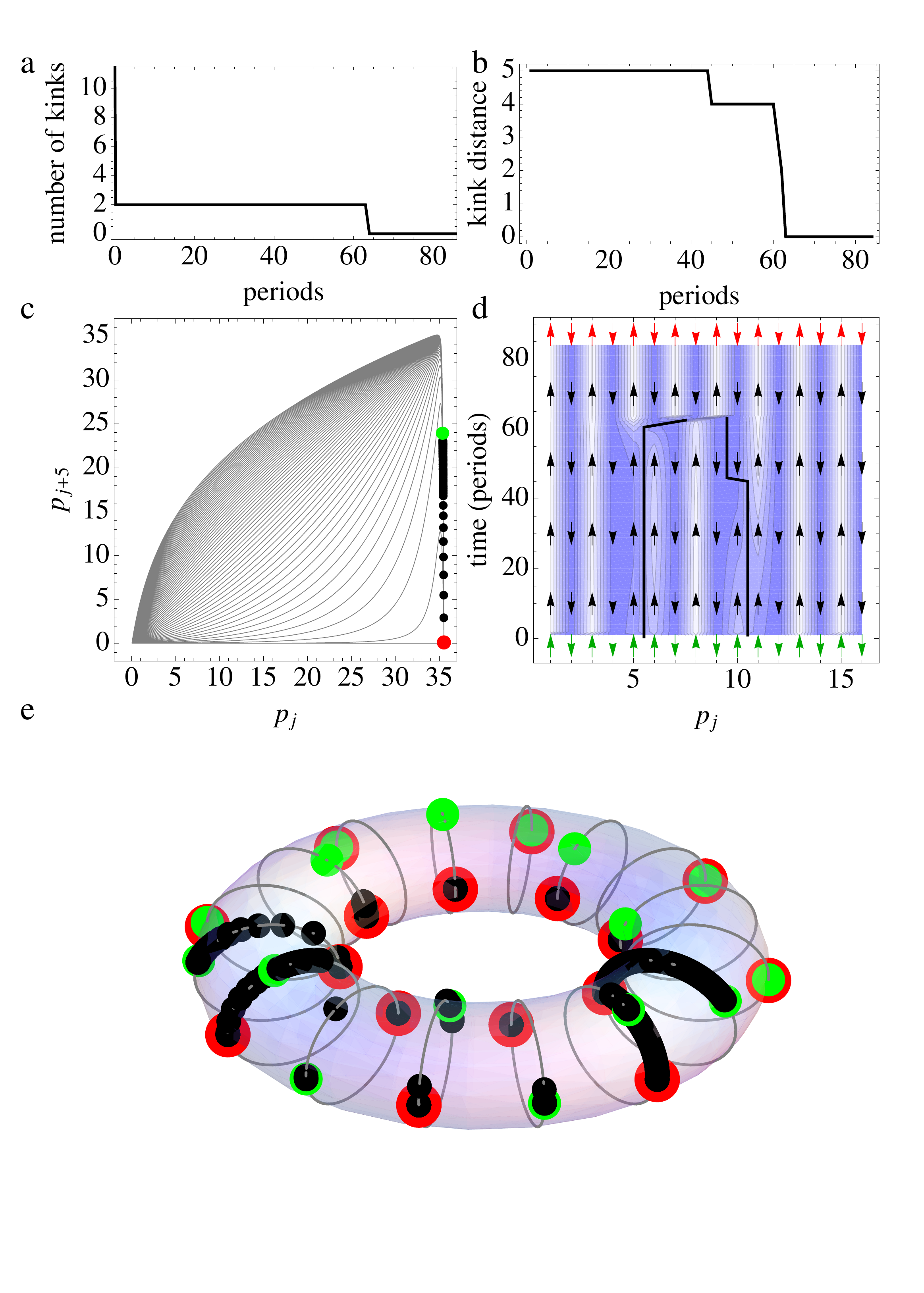}
\caption{ {\bf Slow kink-kink annihilation leads to long oscillatory transients.} 
Dynamics around the UPO arising from the first Hopf bifurcation in the 16-gene ring to
illustrate the long oscillating transient that eventually reaches the stable `up-down' solution. 
{\bf (a)} Starting from a random initial condition, the ring quickly settles into a configuration with two kinks. 
{\bf (b)} The distance between the kinks, $k_d$,  goes from $k_d=5$ to $4$ and quickly through $3, 2$ until the kink-kink annihilation producing a long oscillatory transient of $\sim 80$ periods.
{\bf (c)} Phase plot ($p_1$ versus $p_5$) from the initial condition (light green) to the final point at the stable up-down solution (red).  The black dots correspond to Poincar\'e crossings at equidistant time intervals (every period) and illustrate the very slow initial drift
which only accelerates rapidly at the end of the transient, when the unravelling of the solution takes place. Note also the uniform spreading of the trajectories on the surface of the phase plot, an indication of the reduced dimensionality of the transient dynamics.
{\bf (d)}. Space-time diagram of the transient shows the slow approach of the kinks and their annihilation. The kink traces are indicated by gray lines.
{\bf (e)} Poincar\'e sections in gene space. The value of each protein at the Poincar\'e section is represented as a phasor $\exp{i 2 \pi \phi_j}$ where  $\phi_j=p_j/p_{max}$. The kink annihilation event in this representation corresponds to the unravelling of a "twist", which is a global event involving several ring elements and therefore occurring on very slow time scales. 
}
\label{fig:kinkDyn}
\end{figure}

 Even though even rings act as bistable switches at steady state, the UPOs emanating from the first Hopf in even rings have a strong influence on their observable dynamics. Often, when starting from any random initial condition, even rings of moderate length get trapped into long-lived oscillating transients before they eventually approach the stable `up-down' solution. These observed transients in the numerics can be traced to dynamics around the first unstable UPO, a solution with two kinks in the ring.
A typical transient will quickly settle down into a configuration with two kinks separated by a distance that depends on the initial condition. The two kinks then propagate along the ring for a large number of periods, slowly approaching each other, until they
annihilate. The dynamics of this long transient is shown in detail  in Figure~\ref{fig:kinkDyn} from 
different perspectives.  The very slow slip of the kinks is observable not only through the persistence of the inter-kink distance 
during the transient (Fig.~\ref{fig:kinkDyn}b,d) but also in the phase plot and Poincar\'e sections of the dynamics (Fig.~\ref{fig:kinkDyn}c,e). In particular, the structure of the phase plot (Fig.~\ref{fig:kinkDyn}c) indicates the likelihood that the dynamics of the transient is restricted to a reduced manifold, possibly the unstable torus that emerges from a Neimark-Sacker bifurcation (see
Fig.~\ref{fig:geneRing}c).

\subsection*{Observability of the transient dynamics around UPOs: reachability and persistence}

\begin{figure}[h!]
\includegraphics[width=0.5\textwidth]{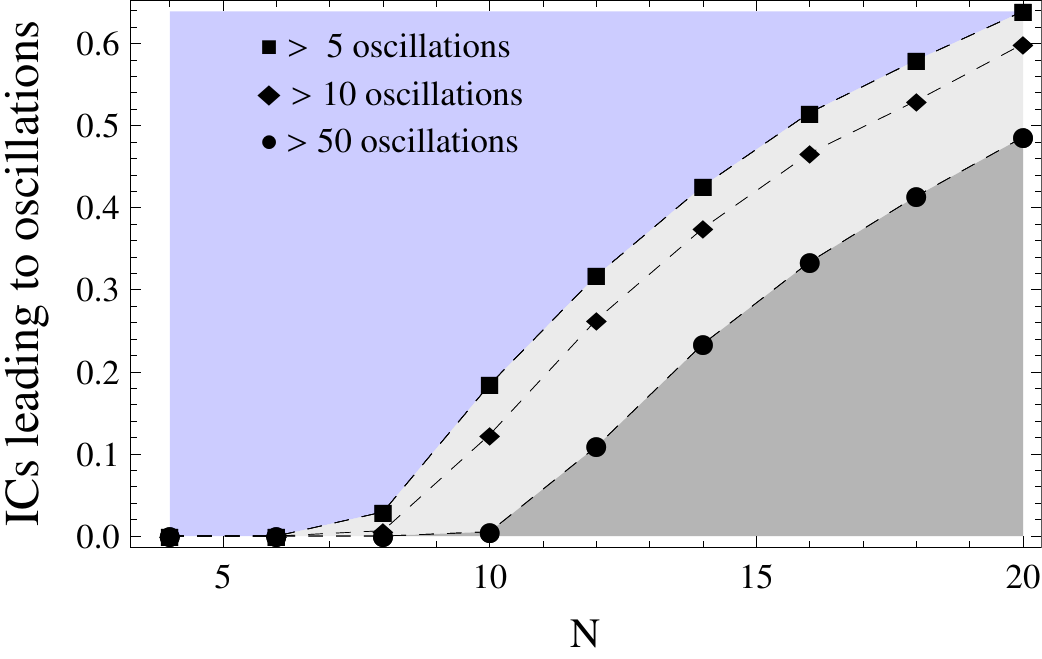}
      \caption{{\bf Likelihood of the observability of long transients starting from random initial conditions.} The probability of observation of oscillating transients with at least 5, 10,  and 50 oscillations is obtained by numerical sampling of the space of initial conditions. The hypercube of initial conditions defined as $([0,m_u]\times [0,p_u])^N$ has been sampled $10^4$ times for each ring of length $N$.  The likelihood of observing long oscillatory transients increases for longer rings.}
      \label{UPOProb}
\end{figure}

\begin{figure}[h!]
\begin{center}
 \includegraphics[width=2.5in]{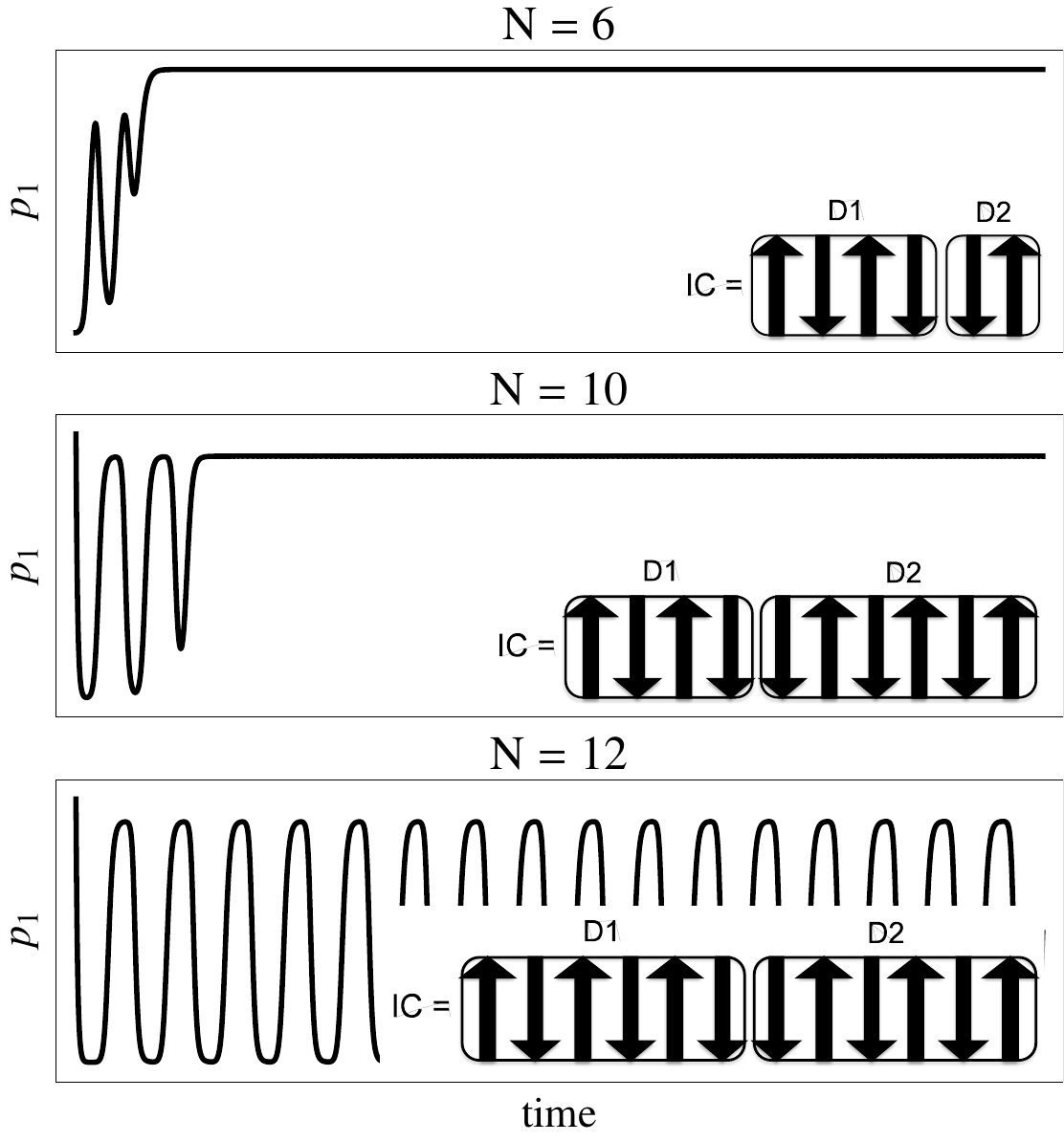}
  \caption{ {\bf Length of oscillating transients in rings of different sizes starting from comparable initial conditions. } The same set of constants has been used for each ring. The insets show the initial conditions used corresponding to idealized `up-down' solutions with 2 kinks separating domains D1 and D2.}
\label{fig:osclDuration}
\end{center}
\end{figure}

For the transient dynamics described above to be relevant for applications and design it needs to be observable, i.e., it should be reachable from large regions of phase space and it should persist over long time scales as compared to the intrinsic time horizon of the system.  Our numerical simulations indicate that the UPOs that emerge from the first Hopf bifurcation in even rings (corresponding to the two-kink arrangement) possess these two attributes. 
Figure~\ref{UPOProb} shows that the probability of observing long oscillating transients is higher than $50 \%$ for moderately long rings. In fact, all the oscillatory transients reached from random initial conditions correspond to the two-kink traveling solution, so this is the only UPO that is observable.  Note also that, as expected, the longer the ring, the more likely it is to observe these transients since it is easier for the initial condition to settle onto a configuration with two well-separated kinks in the ring. This is also seen in Figure~\ref{fig:osclDuration}, where we show the increasing length of the oscillatory transient from a similar initial condition as we increase the length of the ring. These observations explain why although UPOs with more than two kinks are possible in even rings, they are not observed numerically when starting from random initial conditions. Multi-kink solutions are both combinatorially less probable and are more unlikely to settle onto a configuration with well-separated kinks that can persist as a long oscillating transient. 

\begin{figure}[h!]
\includegraphics[width=0.6\textwidth]{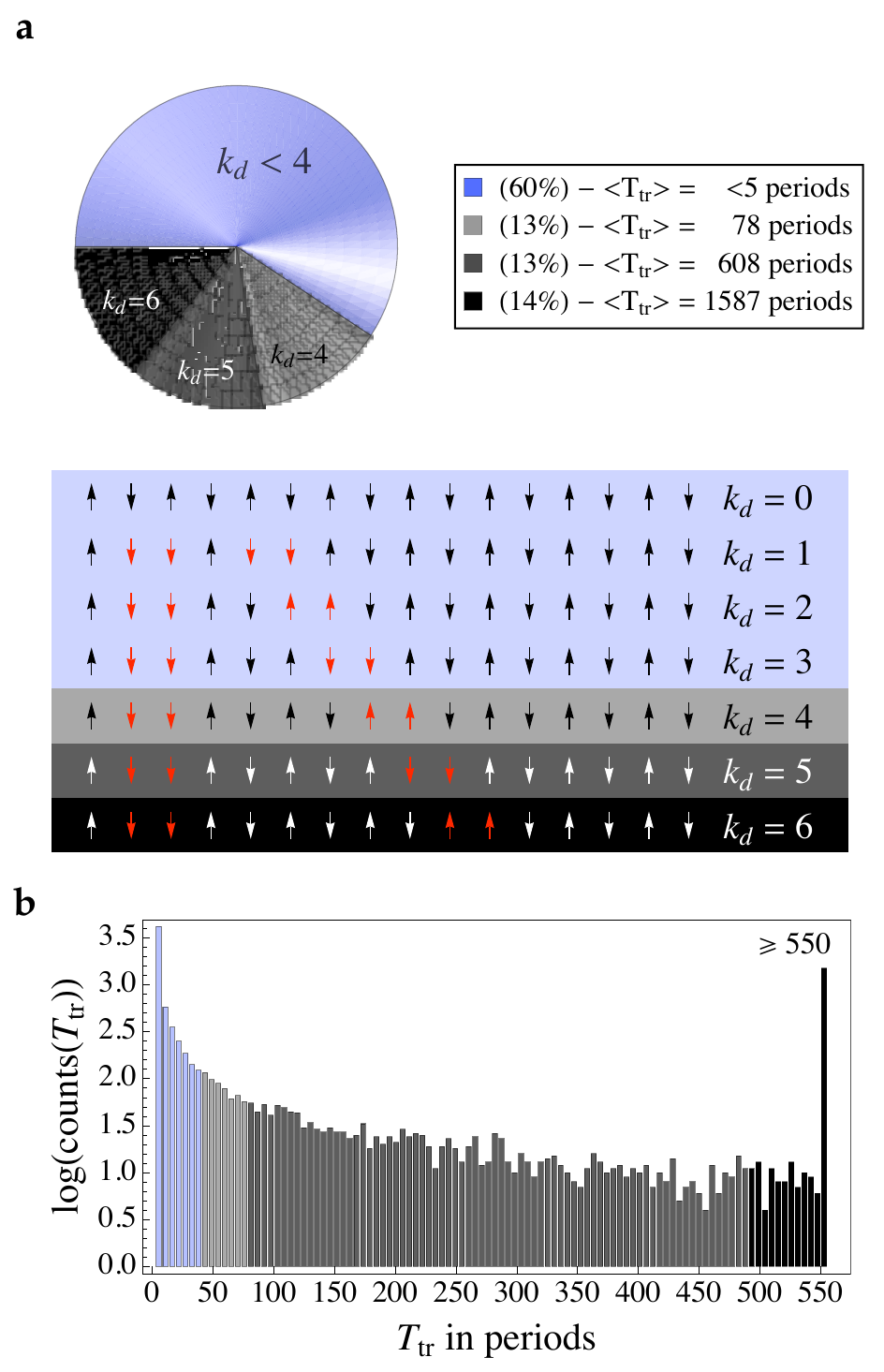}
      \caption{
 {\bf Numerical analysis of oscillating transients starting from random initial conditions for a 16-gene ring.}
{\bf (a)} The pie-chart represents the proportions of observed conformations collected from $10^4$ initial conditions covering the hyper cube $([0,m_u]\times [0,p_u])^N$. For each random initial condition, we measure the distance $k_d$ between the kinks after a burn-in time of five periods. Combinatorially we expected that starting from random initial conditions each configuration should be reachable with comparable probabilities and this is confirmed numerically, as seen in the similar proportions of $k_d=4,5,6$. The proportions for the distances $k_d<4$ are pooled together due to the burn-in time of the simulation.  The table also includes the mean of the distribution of numerically observed lengths of the oscillating transients  for all the two-kink conformations for the 16-gene ring, as represented below. {\bf (b)} Distribution of the length of the transients showing a heavy tail due to the likelihood of observing the very long transients that ensue from the configurations with inter-kink distances $k_d=4,5,6$.
}
      \label{UPOgene16}
\end{figure}

Figure~\ref{UPOgene16} presents a summary of numerical simulations of a 16-gene ring that explore the 
characteristics of the two-kink traveling solution that make it observable. Firstly, random initial conditions settle with high likelihood onto configurations with two kinks separated by a distance $k_d$ that is large enough to lead to a long-lived transient before annihilation.  Figure~\ref{UPOgene16}a shows the numerically observed proportion of initial conditions that settle onto configurations with different inter-kink distances. Note that $40\%$ of the initial conditions lead to configurations with well-separated kinks ($k_d=4, 5, 6$) and the probability of reaching each of these distances is almost equal. 

As explained above, there is a relationship between the duration of the transient and the inter-kink distance in the two-kink configuration.  The further separated the kinks are, the longer the transient lasts, as shown in  Figure~\ref{UPOgene16}a, where we report the mean  \textless$T_{tr}$\textgreater of the distribution of oscillating transient lengths $T_{tr}$.  Note the extremely large average of the transients ensuing from the symmetric (or near symmetric) two-kink arrangements with $k_d=5,6$. These long transients are responsible for the heavy tail of the probability distribution of the period lengths presented in Figure~\ref{UPOgene16}b, which shows that the probability of observing 
a very long transient that oscillates for more than 550 periods is of comparable order to observing the stable `up-down' fixed point directly.

%

\section{Comparison to other unidirectionally coupled rings: flux-gate magnetometers}
\label{magnetoDuffingComp}
There are several systems with a directed ring topology where UPOs play a significant role in the dynamics.
A recent example is the study of UPOs in relation to the emergence of chaotic behavior in rings of Duffing oscillators \cite{Perlikowski10}. Closest to the dynamics of the gene system studied here are the rings of unidirectionally coupled magnetic cores~\cite{Bulsara04} that have been not only studied numerically but also implemented experimentally by Bulsara \textit{et al. }. In the antiferromagnetic case, the system of coupled magnetic cores is described by the following ODEs~\cite{Bulsara04}:
\begin{equation}
\dot{x}_j = - x_j + \tanh \left(c\, \left (x_j - x_{j+1} \right)\right),
\end{equation} 
which falls under the framework of Eq.~(\ref{eq:vectorsystem}).  

The bifurcations branching from the unstable symmetric solution follow from a Jacobian of the same form as that in Eq.~(\ref{eq:jacobian}). Hence the UPO characteristics follow the same pattern as those of the gene system. For even rings, the system shows symmetry breaking when $c$ is increased giving rise first to the two degenerate dimerized solutions followed by the emergence of UPOs from the now unstable symmetric branch. For odd rings, there is also a similar cascade of Hopf bifurcations. The maximal number of Hopf bifurcations due to ring symmetry is $(N/2 -1)$ for even-numbered rings and $(N-1)/2$ for odd-numbered rings, as is the case for the generalized repressilator model. 

Such rings of magnetic cores have been implemented as real-world devices and the odd-numbered 
versions are used as sensitive magnetic sensors. Long even-numbered rings have also been built  
and their dynamics has been explored under noise perturbations and under random initial conditions~\cite{Bulsara06}. It turns out that long oscillatory transients feature heavily in long even-numbered rings, even if the steady state is quiescent~\cite{Bulsara06}.

\section{Discussion}

The relevant dynamics of noisy systems with a limited horizon of observability is not always well defined by their long-term attractors as given by stability analysis. In such systems, long-lived transients can {\it de facto} dominate the observable behavior. Examples of such behavior are prevalent in a variety of engineering, physical and biological systems~\cite{Bamieh04, Trefethen05, Strelkowa10}.  
In this paper, we have analyzed the dynamics of the generalized repressilator model and have characterized the existence of long-lived oscillatory transients in even rings---systems that would otherwise be expected to behave as bistable switches based on strict stability analysis. These transients correspond to slowly decaying dynamics around particular unstable periodic orbits (UPOs) that emerge as part of a cascade of Hopf bifurcations. The UPOs have spatio-temporal symmetries that can be understood in terms of tilings and propagating kinks, corresponding to the moving walls separating different domains of the stable dimerized solution. In particular, the most probable long-lived observable transients correspond to propagation of kink pairs along the ring. We showed numerically that in the presence of such UPOs the dynamics of the system is funneled to its vicinity onto a constrained manifold from which it is difficult to escape, thus leading to observable very long-lived transient oscillations of hundreds of periods when starting from random initial conditions. The likelihood of such behavior can be understood from the fact that when a pair of kinks is well separated, they will propagate along the ring for a long time until they conflate to restore the fixed point solution. This process results in a heavy-tailed distribution of the duration of oscillatory transients.

The gene rings analyzed here present analogies with other systems from classical physics, such as domain wall propagation in antiferromagnetic media or kink propagation in dimerized polymer chains, 
and is mathematically connected with classical models of nonlinear wave propagation in one-dimensional discrete lattices.
However, most of previous models are based on undirected (bidirectional) coupling.  Directed ring topologies with a coupling function that is bounded, monotonic and with a strong enough nonlinearity can lead to the emergence of UPOs with the right values of the coupling strength~\cite{Golubitsky88, Smith87}. These UPOs can then potentially lead to observable long-lived oscillating transients. Recent studies have shown the importance of UPOs in similar directed ring systems, both as originators of transients in the route to chaos in rings of Duffing oscillators~\cite{Perlikowski10} and in the emergence of long-lived oscillatory transients in experimental even-numbered rings of magnetic cores~\cite{Bulsara06}. The linear stability analysis and the cascade of UPOs is similar in all those systems and follows from fundamental symmetry arguments~\cite{Golubitsky88}. However, the reachability and persistence of the
transient dynamics depends on the particular form of the coupling function and reflects global features of the flow in phase space that funnel the dynamics around the oscillatory behavior. Based on our interpretation in terms of kink propagation, it is likely that increasing the nonlinearity of the coupling function (while keeping it bounded and monotonic) will increase the likelihood of localized excitations. Hence, the existence of long-lived transients based on the propagation 
of kink pairs will become more likely. In the case of genetic circuits, this can be achieved by increasing the Hill coefficient $h$ of the repressor. This can be done by the selection of particular gene promoters based on biochemical mechanisms with higher cooperativity.

The design of topologically constrained systems is widely used in classical engineering and applied physics as an aid to controllable dynamics. The generic properties of such systems, which follow from fundamental results from group theory and dynamical systems~\cite{Golubitsky88}, provides a good chance that the key underlying properties could survive under hardly predictable and classifiable biological environments.
In this case, the symmetry of the ring arrangement leads to a strongly symmetric bifurcation analysis and also informs our understanding of the spatio-temporal solutions in terms of tilings and domain walls. 
However these topological considerations require that the elements in the ring be strictly identical. In real systems this condition is not fulfilled and the question remains if UPOs are dynamically observable when the individual elements
are dissimilar, both in numerical simulations and in real-world built systems. The implementation of the ring of magnetic cores by \citet{Bulsara06} suggests that UPOs do feature in real-world applications. Our numerical analysis in \citet{Strelkowa10} showed that UPOs are also observable in gene repressilator rings when the genes are not identical. They also feature heavily in the dynamics when the system is fundamentally stochastic, i.e., in the equivalent Markov process (Master equation). 
Such robustness allowed us to introduce the concept of a switchable synthetic genetic oscillator, operating under uncertainty and noise, based on the feedback operation around the UPOs associated with the long-lived transients.
These examples and others indicate that the conceptual design based on simplified designs can still lead  to a fundamental understanding, although system-specific characterizations in the presence of noise and uncertainty are essential when applications to be considered. Such a perspective is of a potential value to provide ideas for transferable design unifying several disciplines, ranging from electrical engineering to
synthetic biology.


\begin{acknowledgments}
N.S. gratefully acknowledges support from the Wellcome Trust through grant 080711/Z/06/Z. We thank Andreas Ipsen, Kim Parker and Guy-Bart Stan for discussions.
\end{acknowledgments}

\appendix

\section{Analytical conditions for the number of Hopf bifurcations}
Properties of the nonlinear coupling functions and the length of the ring $N$ all determine how many Hopf bifurcations and which UPOs are expected for a given parameter set $\{c_1,c_2,c_3,c_4\}$ of the generalized repressilator model. 

Consider the case with the Hill repressor function (Eq.~\ref{eq:Hill}). Then the 
characteristic polynomial~(\ref{eq:characteristic_polynomial}) becomes
\begin{equation}
\lambda_\ell^2 + (c_2 +c_4) \lambda_\ell + c_2c_4\left(1+\alpha \exp(i \, \beta_\ell(N)) \right)=0
\end{equation}
where $\alpha = \frac{h p^{h+1}_m}{c}$ and $\beta_\ell=\frac{2 \pi \ell}{N}$ with  $\ell=\{0,..,N-1\}$. 

The necessary condition for the Hopf bifurcation to occur is purely imaginary eigenvalues of the Jacobian. For a complex  quadratic polynomial $p(z)$
\begin{equation}
(a+iA) z^2 +2(b+iB)z + (c+iC)=0
\end{equation}
the necessary condition for purely imaginary roots is~\cite{Hardy}
\begin{equation}
C^2-4bBC-4b^2c=0.
\end{equation}
In our case, $A=B=0$, $b=(c_2+c_4)/2$, $c=c_2c_4\left(1 + \alpha \, \cos \beta_\ell \right)$ and 
$C= c_2c_4\alpha \, \sin \beta_\ell$, which leads to the condition:
\begin{eqnarray}
\alpha^2 c_2c_4(1 -\cos^2 \beta_\ell)-(c_2 +c_4)^2(1+\alpha \, \cos \beta_\ell)=0. \nonumber\\
\label{eq:quadraticHopfCond}
\end{eqnarray}
This equation determines which values of $\beta_\ell(N)$ can satisfy the condition for Hopf bifurcation for a particular parameter set. 

Taking into account the symmetry of the cosine function and the fact that $\alpha \textgreater 0$, we recover the classical result~\cite{Golubitsky88} that the maximum number of solutions permitted by the ring symmetry is
\begin{eqnarray}
\begin{cases} 
(N-1)/2 \quad & \text{for odd rings} \\ 
(N/2-1) \quad &\text{for even rings}.
\end{cases}
\end{eqnarray}  
However, depending on the parameter set some of those solutions might not be observed in numerical simulations.
In our particular case, with Hill coefficient $h=2$ and for the biologically relevant parameters~\cite{Elowitz00}: $c_2 = 0.12\,\text{min}^{-1}, c_3 = 0.16\,\text{min}^{-1}, c_4 = 0.06\,\text{min}^{-1}$ and bifurcation parameter $c_1 \in (0, 3]$, we get that
\begin{eqnarray} 
\begin{cases}
\small{\left \lfloor\frac{1}{2}\frac{(N+1)}{2}\right\rfloor} \quad & \text{for odd rings} \\
\vspace*{-.12in} \\
 \small{\left \lfloor \frac{1}{2} \left(\frac{N}{2}-1\right)\right\rfloor} \quad &\text{for even rings}.  
\end{cases}
\label{eq:NumberHopfs}
\end{eqnarray}
This set of solutions satisfies the bifurcation conditions for $\beta_l(N) \in \left (\frac{3}{2} \pi, 2 \pi \right)$ 
and leads to the existence of UPOs in even rings with $N \geq 6$ as can be seen by using the
bound
\begin{eqnarray}
0 < \alpha = \frac{h \, p_m^{h+1}}{c} = h \left (1 - \frac{p_m}{c} \right ) < h \nonumber.
\label{eq:inequality}
\end{eqnarray}
Interestingly, for the same set of parameters $\{c_i\}$ but with a higher Hill coefficient $h=4$, 
UPOs exist in smaller even rings and the criterion~(\ref{eq:quadraticHopfCond})  
is fulfilled for $ N \geq 4$.

\end{document}